\documentstyle[12pt,epsfig,amssymb]{article}
\begin{document}
\setlength{\parskip}{0.45cm}
\setlength{\baselineskip}{0.75cm}
\begin{titlepage}
\begin{flushright}
CERN-TH/98-377 \\ 
{\tt hep-ph/9812231} \\
\end{flushright}
\vspace{1.5cm}
\begin{center}
\Large
{\bf On the photon constituency of protons}\\
\vspace{0.5cm}
\large
A.\ De R\'{u}jula and W.\ Vogelsang \\

\vspace*{0.5cm}
\normalsize
{\it Theoretical Physics Division, CERN, CH-1211 Geneva 23, Switzerland}

\vspace{2.0cm}
%
\large
{\bf Abstract} \\
\end{center}
\vspace*{-0.3cm}
We argue that existing measurements of $ep$ collisions
at HERA --in which an energetic photon is made via a 
QED `Compton' subprocess-- 
can provide rather detailed
information on the photonic parton density of the proton.
This function and its deviations from Bjorken scaling should be
measurable, allowing for an interesting test of the theory. The photonic
distribution function and its gluonic counterpart should
show a strikingly different evolution with momentum scale.

\vspace{5.0cm}

\noindent CERN-TH/98-377 \\
November 1998
\end{titlepage}
\newpage
\normalsize
%

\section{Introduction}
\noindent
The $ep$ collisions in which there is
an energetic photon in the final state can be classified according
to their distinctive kinematical features.
We are concerned in this note with 
the Compton subprocess $e\gamma \rightarrow e\gamma$, where 
the initial photon is coupled to the proton and is (almost) 
on-shell.
Progress at HERA \cite{h1gam} and the subsequent prospects
for improved measurements induce us to revisit, update and extend
earlier theoretical work on this subject \cite{kniehl}--\cite{gsv}.

Figure~1 shows the lowest-order `Compton' Feynman diagrams,
 along with our notation for the kinematical variables.
At the HERA collider,
events of this type have a particularly clean and distinctive signature, as 
there are only an electron\footnote{We call both electrons
and positrons `electrons'.} and a photon in the final state, with 
little or even no observable hadronic activity. In addition, to lowest 
order in perturbation theory, the (large) transverse momenta of the 
electron and the photon approximately balance each other.
Conversely, applying these distinctive features as selection criteria
allows one to extract, to the leading log approximation, Compton events
from the ensemble of radiative corrections \cite{radiative}.
\begin{figure}[hbt]
\begin{center}
\vspace*{-1.6cm}
\hspace*{-1cm}
\epsfig{file=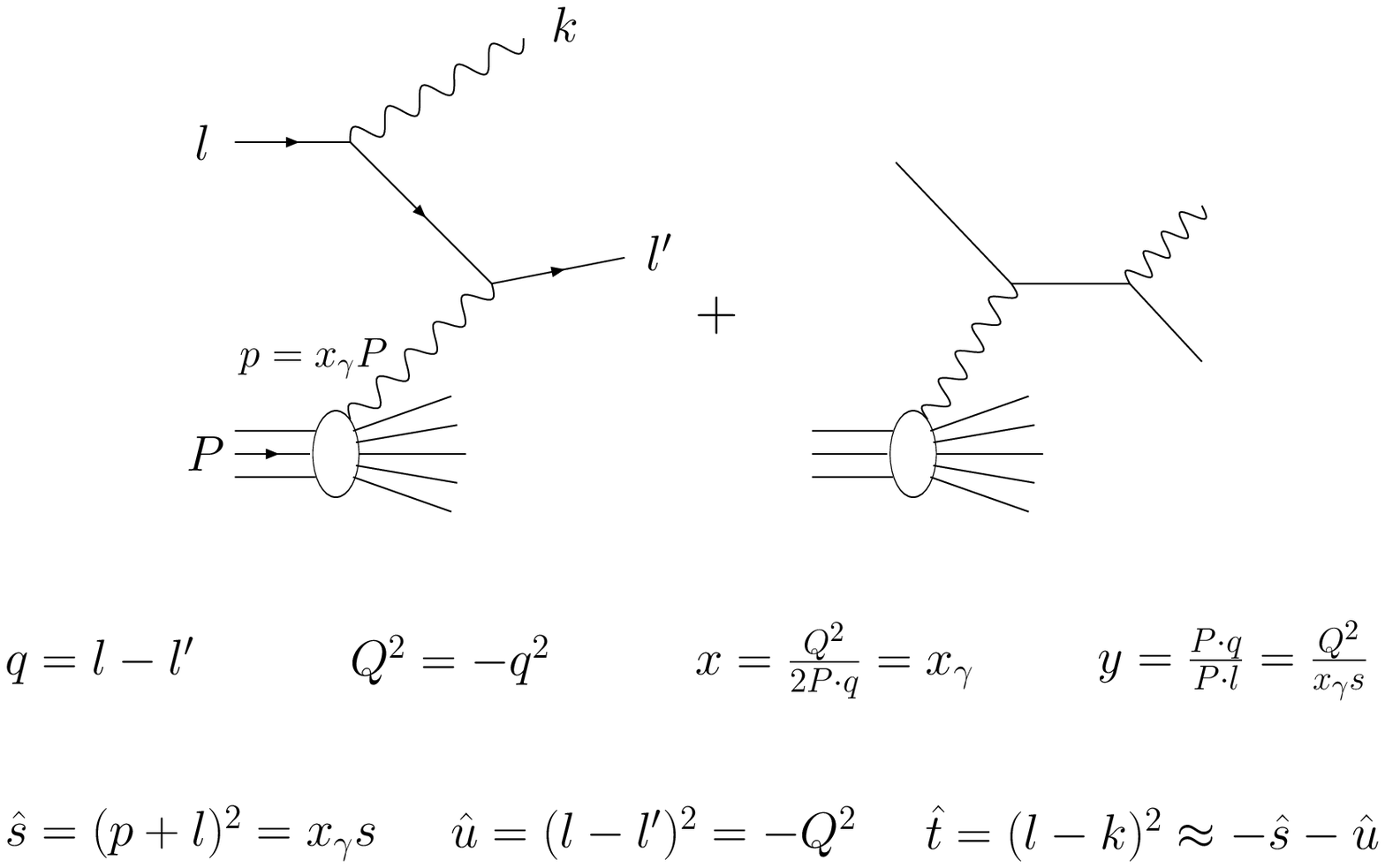,width=18cm}
\vspace*{-14.6cm}
\caption{\sf Lowest-order 
Feynman diagrams for Compton scattering in $ep$ collisions.}
\vspace*{-0.5cm}
\end{center}
\end{figure}

We are interested in the $ep\to e\gamma\, X$ Compton process at
relatively large momentum scales $Q^2$; we refer to it
as Deep Inelastic Compton Scattering
(DICS). The experimental study of DICS at HERA offers the 
unique possibility of measuring the photon--parton content of the proton
and the corresponding longitudinal-momentum distribution 
or `structure' function 
$\gamma(x,Q^2)$.
In the `parton model' approximation in which 
the initial photon is on-shell and collinear with the proton, 
the relation between the
DICS differential cross section and $\gamma(x,Q^2)$ is~\cite{blum2}:
\begin{equation} \label{sigma}
\frac{d^2 \sigma (s,x,y)}{dx\, dQ^2}  = \int_x^1 \frac{dz}{z} \;
\frac{d^2 \hat{\sigma}^{e\gamma\rightarrow e\gamma}(x s,x/z,y )
}{d(x/z)\,dQ^2} \;
 \gamma (z,Q^2) \; , 
\end{equation}
where
\begin{equation} \label{sigma1}
\frac{d^2 \hat{\sigma}^{e\gamma\rightarrow e\gamma}(\hat{s},\hat{x},y)}
{d\hat{x}\, dQ^2}
 =
\frac{2 \pi \alpha_{em}^2}{\hat{s}^2}\; \frac{1+(1-y)^2}{1-y}\; 
\delta (1-\hat{x}) \; .
\end{equation}

A measurement of $\gamma(x,Q^2)$ would acquire 
particular interest when compared with that of
its non-Abelian counterpart: the gluon distribution 
function $g(x,Q^2)$.
The predictable $Q^2$ evolution of these functions ought to be very different,
their difference representing a very clean test of the non-Abelian nature
of the gluon. More specifically, the self-coupling of gluons makes 
$g(x,Q^2)$ evolve with $Q^2$, at small $x$, much faster than
$\gamma(x,Q^2)$. The evolution of $\gamma(x,Q^2)$ with $Q^2$
is by itself an interesting test of the QCD-based parton picture, which we
argue can be performed at HERA with the existing or soon to be gathered 
statistics.

One can in principle, and sometimes in practice, 
distinguish two types of contributions to DICS: `pseudo-elastic' 
$ep\rightarrow e\gamma p$, and `inelastic'   
$ep\rightarrow e\gamma X$ with $X\neq p$.
From a parton model point of view, in which
the struck photon in Fig.~1 is viewed as a proton constituent,
the distinction between pseudo-elastic and inelastic subprocesses
is very artificial. This is best understood by comparing conventional
deep inelastic scattering with the Compton process, as we
do in Fig.~2. Deep inelastic scattering, Fig.~2a, results in two final-state
jets: the `current' or struck quark jet and the `spectator' jet
of target fragments.  The invariant
masses of these jets (to the extent that they can be 
ascertained without ambiguity) are small; 
the unavoidable colour reconnection between
the struck quark and the target fragments is in general a soft process. 
At high $Q^2$
a successful recombination leading to an elastic event $ep\to ep$ is a rare
ocurrence. Similarly, the photon and the target fragments in the Compton
process of Fig.~2b would even more
rarely recombine into a proton, to result
in a literally elastic scattering event $ep\to ep$. There is nothing very
special about the `pseudo-elastic' $ep\rightarrow e\gamma p$ channel,
which ought to be quite common, since the invariant
mass distribution of the target fragments will, as in deep inelastic
scattering, tend to be low. In DICS the total invariant mass of the 
proton's fragments (the hadrons as well as the outgoing photon)
will, again as in deep inelastic scattering, be large in general.

\begin{figure}[hbt]
\begin{center}
\hspace*{-1.8cm}
\vspace*{12cm}
\epsfig{file=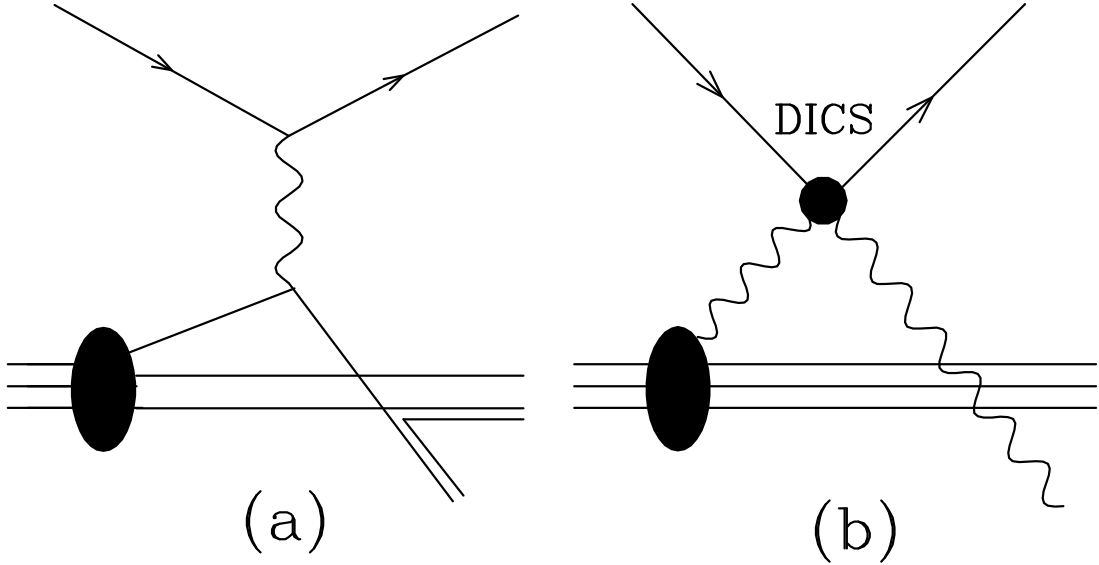,width=13cm}
\vspace*{-5cm}
\caption{\sf Diagrams for deep inelastic conventional and `Compton'
scattering.}
\vspace*{-0.5cm}
\end{center}
\end{figure}

The pseudo-elastic contribution to DICS can be easily worked out in
terms of measured proton form factors \cite{kniehl}. In the current
state of our understanding of QCD, it is simply impossible to predict
the contribution of final hadronic states of invariant mass $W_H>m_p$.
To study the feasibility of measuring  $\gamma(x,Q^2)$ and its $Q^2$ evolution
in DICS we must, as
other authors \cite{blum1}--\cite{gsv} have done before us, proceed to make 
a series of conjectures. Mercifully, these conjectures
 are not relevant to
the $Q^2$ evolution, which is a solid prediction of the standard model.

\section{Pseudo-elastic DICS}

The contribution of the pseudo-elastic channel 
$ep\to ep\gamma$ to $\gamma(x,Q^2)$ can be explicitly
written down in terms of the proton's elastic form factors \cite{kniehl}.
The usual `electric' and `magnetic' form factors are empirically well
fit as dipoles:
\begin{equation} 
G_E (t) \simeq\frac{1}{[1-t/(0.71 \, {\mbox{GeV}}^2)]^2} \; , \; \; \; 
G_M (t) \simeq 2.79\; G_E (t) \; .
\end{equation}
Define the quantities:
\begin{equation}
H_1(t) \equiv \frac{G_E^2 (t)-(t/4m_p^2) G_M^2(t)}{1-t/4m_p^2} \; , \; \; \; 
H_2(t) \equiv G_M^2 (t) \; ,
\end{equation}
to express the
pseudo-elastic contribution to $\gamma(x,Q^2)$ as:
\begin{equation} \label{gamel}
\gamma_{el} (x) = -\frac{\alpha_{em}}{2\pi} x \int_{t_1}^{t_2}
\frac{dt}{t} \Bigg\{ 2 \Bigg[ \frac{1}{x} \left( \frac{1}{x}-1 \right) 
+\frac{m_p^2}{t} \Bigg] H_1 (t) + H_2 (t) \Bigg\} \; , 
\label{elastic}
\end{equation}
where the limits on the virtual photon mass $t$ are:
\begin{equation}
t_{1,2} = 2 m_p^2 -\frac{1}{2s} \left[ (s+m_p^2) (s\, [1-x] +m_p^2)
\pm (s-m_p^2) \sqrt{(s\, [1-x]+m_p^2)^2-4 s m_p^2} \, \, \right] \; .
\end{equation}

The tempting analogy between photons and gluons as
partons in the proton breaks down at various points.
One of them is that colour confinement and conservation
preclude the existence of a strict coloured analogue to 
a quasi-elastic channel: there are no QCD elastic form
factors. A neutral and spinless hadron,
such as a $K_L$, would have a very small quasi-elastic
contribution to its photonic structure function $\gamma_K(x,Q^2)$,
satisfying in this respect a QED/QCD analogy more 
closely than a proton does.
But the proton carries a long-range photon field and 
$\gamma_{el} (x)$ is an important contribution 
to its $\gamma(x,Q^2)$. Another departure from the
photon/gluon analogy is that, in DICS, it
is justified to work to leading order in $\alpha_{em}$,
and to this order
the quantity $\gamma_{el} (x)$ of Eq.~(\ref{elastic}) is independent
of $Q^2$. 

At an $ep$ collider, a separation of elastic and 
inelastic Compton events is difficult but possible.
A fraction of elastic events could be tagged by
the very forward `diffraction' 
detectors. Inelastic events with
proton fragments at sufficiently large $p_T$
could be caught by larger-angle detectors.
Although the elastic contribution can be ascertained
with accuracy, we see no particular
interest in singling it out experimentally except, perhaps, 
as a calibration and/or luminosity check \cite{h1gam}.

\section{Inelastic DICS}

In principle one could build-up the complete
function $\gamma(x,Q^2)$ by adding to the quasi-elastic contribution
all resonant \cite{cour} and non-resonant
final hadronic states, and their interferences. 
Alternatively one could directly guess, one way or 
another, an inclusive or `continuous'
non-elastic part of $\gamma(x,Q^2)$. If this guess
is based  on a parton
picture wherein the photon `constituent' is emitted by one
of the quarks in the proton \cite{blum1,blum2,gsv}, 
the addition of the continuous
and resonant contributions may be double-counting, as it
would certainly be in $e^+e^-$ annihilation \cite{ee}
and arguably be in deep-inelastic scattering \cite{dis}.
In these latter cases the data provide the decisive proof
that there is a `duality' between continuous and resonant
contributions: adding them is double-counting.
 In DICS, the information on $\gamma(x,Q^2)$
is at the moment too sparse to help decide on this interesting
question.

Having to make a guess, we choose to make one that
is correct in that aspect of $\gamma(x,Q^2)$ for which the
QCD prediction is unequivocal: the $Q^2$ evolution.
Two diagrams for the contribution of quark--gluon scattering
to high-$p_T$ photon production are shown in Fig. 3. Notice
that, with the substitution of gluons for photons and
of a parton--quark for an electron,
they are identical to the Compton diagrams of
Fig.~1. This implies that
DICS is `factorizable' in the same sense as `Drell--Yan'
scattering or the production of high-$p_T$ photons or jets in hadronic
collisions \cite{soper}. In turn, this means 
that $\gamma(x,Q^2)$ satisfies a `QED evolution' equation \cite{Russians}.
To lowest order  in $\alpha_{em}$ and $\alpha_s$~\cite{gsv}, and
in an obvious notation\footnote{The $O(\alpha_{em})$
modifications of the quark and gluon evolutions induced
by a non-vanishing $\gamma(x,Q^2)$, affect the evolution
of the latter only at $O(\alpha_{em}^2)$.}:
\begin{equation} \label{evol}
\frac{d\gamma (x,Q^2)}{d \ln Q^2} = \frac{\alpha_{em}}{2 \pi}
\int_x^1 \frac{dy}{y} \sum_q e_q^2\, P_{Aq} \left( \frac{x}{y} \right)
\Big[ q(y,Q^2)+\bar{q} (y,Q^2) \Big] \; , 
\end{equation}
where  the quark-to-gauge-boson 
splitting function is:
\begin{equation}
P_{Aq}(z)=\frac{1+(1-z)^2}{z} \; .
\end{equation}
A function $\gamma(x,Q^2)$ satisfying Eq.~(\ref{evol}) is
automatically
correct to the leading logarithmic order in QED. This is
also the order to which it is permissible to single out
Compton scattering from the other radiative processes
in $ep$ collisions.

We have chosen the lepton-to-lepton squared momentum transfer
$Q^2$ as the scale in Eq.~(\ref{evol}).
A scale  $p^2_T(e,\gamma)$ or either of the quantities
$\hat s$, $-\hat t$ defined in Fig.~1 would be equally reasonable. 
To leading order of perturbation theory there is no way to favour
any particular choice.

\begin{figure}[hbt]
\begin{center}
\hspace*{-0.3cm}
\epsfig{file=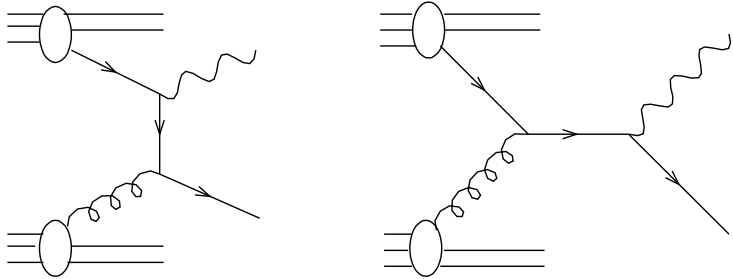,width=10cm}
\caption{\sf Quark--gluon scattering in high-$p_T$ $\gamma$
production in $pp$ collisions.}
\end{center}
\end{figure}

The non-Abelian generalization of Eq.~(\ref{evol}) is the gluon-evolution
equation~\cite{ap}:
\begin{equation} \label{evolg}
\frac{dg (x,Q^2)}{d \ln Q^2} ={\alpha_s (Q^2)\over 2 \pi}
\int_x^1 \frac{dy}{y} \Bigg[ \sum_q {4\over 3} P_{Aq} \left( \frac{x}{y} 
\right)
\Big[ q(y,Q^2)+\bar{q} (y,Q^2) \Big] + P_{gg} \left( \frac{x}{y} \right) 
g (y,Q^2) \Bigg]  , 
\end{equation}
where the (lowest-order) gluon-to-gluon splitting function 
$P_{gg}$ is given in~\cite{ap}. 
An interesting challenge to experiment is to observe the 
different $Q^2$ evolutions of $\gamma$ and $g$, as predicted
in Eqs.(\ref{evol}) and (\ref{evolg}).

To illustrate the experimental feasibility of a measurement of
$\gamma(x,Q^2)$ and its evolution, we must guess that
function explicitly. For the guess to be consistent with QCD, 
it must be compatible with Eq.~(\ref{evol}). We follow~\cite{gsv} in
writing $\gamma=\gamma_{el}+\gamma_{in}$ and figuring
out $\gamma_{in}(x,Q^2)$ by evolving an input 
$\gamma_{in}(x,Q_0^2)$ with the use of Eq.~(\ref{evol}).
We also follow~\cite{gsv} in assuming that $\gamma_{in}$ vanishes
at a  low $Q_0=0.5\,{\rm GeV}$; it builds up at $Q^2>Q_0^2$
via evolution. This is admittedly as arbitrary
for a photon constituency as it would be for its gluon counterpart;
the proton being a bound state of charged coloured objects, there
is no reason for it to be `made' of only quarks at any scale, even
at leading twist and leading order of perturbation theory. 
We differ from~\cite{gsv} in our use of 
a more recent set~\cite{grv} of parton densities\footnote{This
leads to an increase of the conjectured $\gamma (x,\mu^2)$ at small $x$,
 relative to the result in~\cite{gsv}.}.

We deal with momentum scales comparable to the proton mass and
we should be making target mass-corrections distinguishing
Bjorken's $x$-scaling \cite{bj}  from Nachtmann's $\xi$-scaling
\cite{otto}. As it turns out, $\gamma(x,Q^2)$ is only measurable
at values of $x$ small enough for this distinction not to be
important.
 
\section{Experimental details and expectations}

In~\cite{h1gam}, a first 
measurement of DICS at HERA was presented; at that time 
the main aim was the use of Compton events as an independent 
luminosity monitor. The roughly
400 reported events were insufficient to
study differential distributions and to provide a 
measurement of $\gamma$. Moreover, for many of these events 
it was not possible to tell which of the electromagnetic showers
was the electron and which was the photon. Since then, about two orders 
of magnitude more luminosity have been accumulated. In addition, 
silicon trackers have been installed in the backward region, and charge
identification is now possible with much higher efficiency. 
We show that, given these
improvements, a lot of information on
 $\gamma(x,Q^2)$ can be extracted from existing data.

To match the experimental situation,
we use the following parameters and cuts, the latter akin to the ones 
used for event selection in the quoted H1 analysis~\cite{h1gam}:
\begin{itemize}
\item The beam energies are $E_e=27.5$ GeV and
$E_p=820$ GeV. We slightly over-estimate event rates by using 
the current 
value of ${\cal L}=36.5$/pb for the total luminosity\footnote{Our 
numbers
can be easily rescaled to  
the luminosity collected {after} the installation of the silicon trackers.}
collected by H1.
\item The electron and the photon are both seen by the central 
detector, i.e. they have $0.05$ rad $\leq \theta_{e,\gamma} \leq \pi-0.05$ rad
(as usual, angles are measured with respect to the proton beam direction). 
\item At least one of the electromagnetic clusters has an energy of more than 
8 GeV, and both of them have at least 5 GeV. The latter requirement guarantees 
a high detection efficiency.
\item Other cuts in~\cite{h1gam} are 
automatically satisfied at the level of our leading-order calculation. 
The total visible electromagnetic energy is always larger than  
18 GeV (the final $e+\gamma$ energy equals 
$E_e + x_{\gamma} E_p > 27.5$ GeV). Except in the forward region
there is no additional (hadronic) cluster with more
than 2 GeV energy.
The $e\gamma$ 
acoplanarity angle $\Delta \phi \equiv \pi - |\phi_e -\phi_{\gamma}|$  
is below $45^{\rm o}$.
\end{itemize} 

We also demand that no momentum scale be small: for the quantities
defined in Fig.~1,
$-\hat{t},\,\hat{s},\, Q^2 >1$ GeV$^2$;  for 
the (equal and opposite) transverse momenta of the final-state electron
and photon, $p_T^{e,\gamma}>1$ GeV. 

The first relevant question concerns the number of DICS events, with the
quoted integrated HERA luminosity.
In Fig.~4 we show, in bins of  $\log_{10} x$, 
the event numbers that survive the above selection criteria. Clearly,
they should suffice to measure this $x$-distribution,
a point previously emphasized in \cite{blum1}--\cite{cour}. 
The electron and photon detection efficiency (and presumably the 
ability to distinguish them) is high within the applied cuts.
We shall thus estimate the statistical errors as the square root 
of the event number per bin. To measure $x$, $e/\gamma$ identification
is unnecessary, since $x=x_{\gamma}$ may be 
determined from the requirement that the sum of the  
energies of the two electromagnetic clusters be equal to
the combination $E_e+x_{\gamma} 
E_p$. Contrawise, for the determination of 
$Q^2$, an $e/\gamma$ distinction is indispensable.
\begin{figure}[hbt]
\begin{center}
\hspace*{-0.3cm}
\epsfig{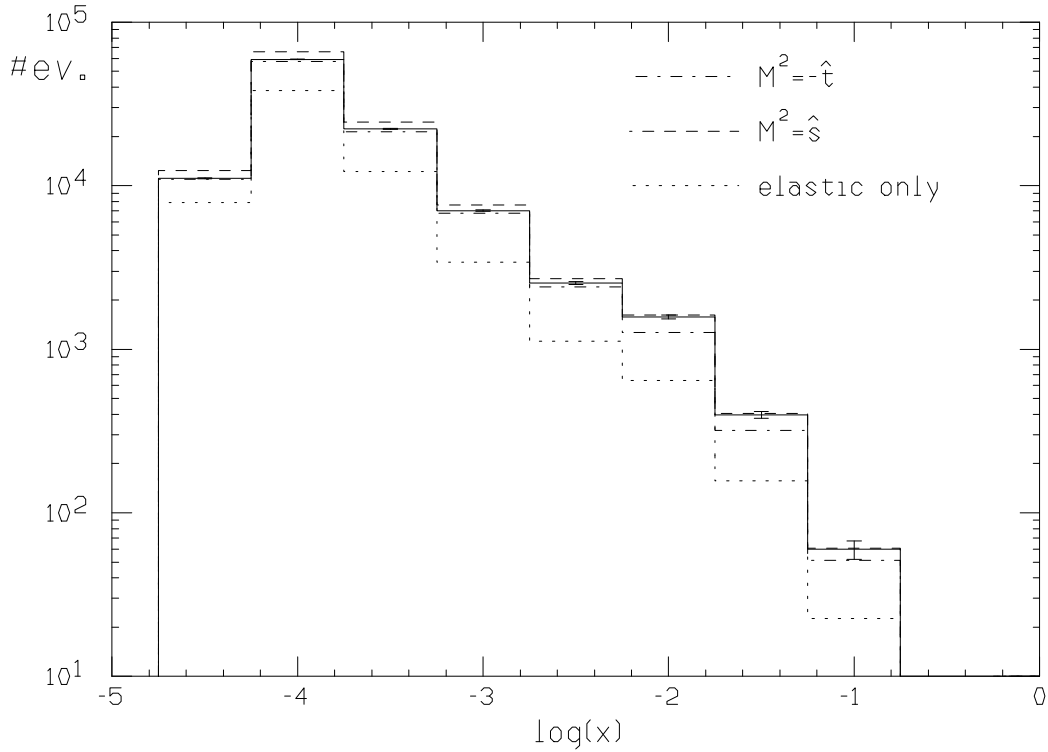}
\caption{\sf Event rates for the DICS at HERA. The cuts
applied are as described in the text. The lowest-$x$ bin contains
all events with $\log_{10} (x) \leq -4.25$.}
\vspace*{-0.5cm}
\end{center}
\end{figure}

In Fig.~4 the dashed (dash-dotted) histogram 
depicts the number of DICS events found  by
integration of Eq.~(\ref{evol}) using a scale 
${\hat{s}}$ (${-\hat{t}}\;$) instead of $Q^2$.
The scale dependence 
is a substantial theoretical uncertainty, but is not devastatingly large. 
The fact that the elastic part is
scale-independent helps make the total theoretical
expectation fairly stable. The dotted histogram in Fig.~4 shows the 
event rates predicted on the basis of quasi-elastic scattering only:
$\gamma=\gamma_{el}$.  
It should be an easy task for experiment to confirm, or infirm, the 
existence of the substantial inelastic contribution that we foretell.

To illustrate the experimental extraction of $\gamma(x,Q^2)$
we translate the information in Fig.~4 into a statement on the 
accuracy of the measurement. To this end,
we evaluate $\gamma (x,Q^2)$ at the 
statistical averages $\langle x\rangle,\langle Q^2\rangle$ determined 
from the event sample used in Fig.~4. We assume that in each bin 
the error in $\gamma$ is only statistical. The result 
for $x\gamma/\alpha_{em}$ as a function of $\log_{10}(x)$ is
shown in Fig.~5.
For comparison, we also show the contribution of $\gamma_{el}$.
In practice a possible method~\cite{unfol}
to translate the data into a measurement of $\gamma$
is based on the bin-by-bin evaluation and subsequent iteration
of the expression:
\begin{equation} \label{meas}
\gamma^{meas} [\langle x\rangle,\langle Q^2\rangle] \equiv
\frac{\# ev., \, \mbox{data}}{\# ev., \, \mbox{MC} [\gamma^{toy}]} \cdot 
\gamma^{toy} [\langle x\rangle,\langle Q^2\rangle ] \; , 
\end{equation}
where  $\# ev., \, \mbox{MC} [\gamma^{toy}]$  is the
expected (theoretical or MonteCarlo) number
of events, based on a trial function $\gamma^{toy}$. 
If $\gamma^{toy}$ is not too far off
the true structure function $\gamma$, the iteration
of Eq.~(\ref{meas}) converges fast. A reasonable 
guess is $\gamma^{toy} \equiv \gamma_{el}$. To see how
well this choice works, we make use of 
it on the right-hand side of Eq.~(\ref{meas}), 
along with our results of Fig.~4 
as the `data sample'. The $x\gamma^{meas}/\alpha_{em}$
 so determined is also shown in Fig.~5.
Even this `zeroth' 
order iteration is accurate.

\begin{figure}[hbt]
\begin{center}
\epsfig{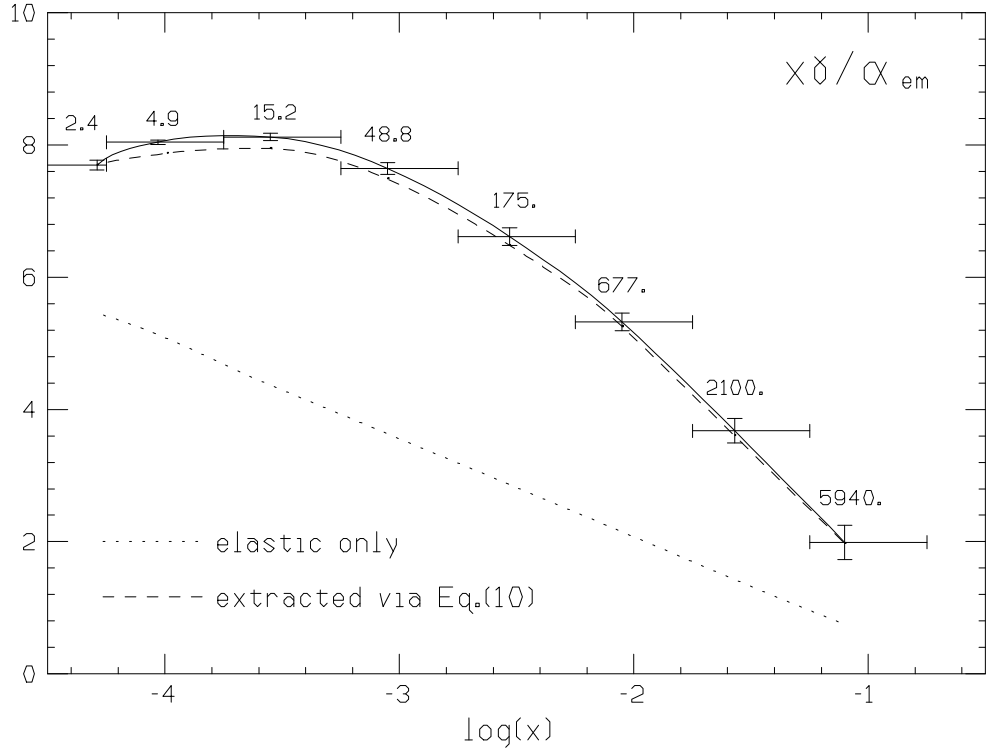}
\caption{\sf Expected statistical accuracy of the determination 
of $x\, \gamma (\langle x\rangle,\langle Q^2\rangle)$.
The numbers indicate the average $\langle Q^2 \rangle$ (in GeV$^2$) for each 
$x$-bin. The dotted line shows the purely elastic spectrum, while the 
dashed one corresponds to $\gamma^{meas}$ `extracted' via Eq.~(\ref{meas})
and as explained in the text.}
\vspace*{-0.5cm}
\end{center}
\end{figure}
To study the detectability of the $Q^2$ dependence of $\gamma$,
we take the sample used for Figs.~4 and 5 and bin it additionally in 
$Q^2$. The result is shown in Fig.~6. There
is an increase of $x \gamma (x,Q^2)/\alpha_{em}$ with $Q^2$ that 
should be observable in most $x$-bins. 
We also display  
the $Q^2$ 
dependence of the leading-order GRV~\cite{grv} gluon density, which satisfies
Eq.~(\ref{evolg}). To facilitate the comparison of the relative
variations of $\gamma$ and $g$, we renormalize $x\, g(x,Q^2)$
in each subplot so as to coincide with $x\gamma (x,Q^2)/\alpha_{em}$ at the 
lowest accessible $Q^2$.
\begin{figure}[p]
\begin{center}
\vspace*{-1cm}
\epsfig{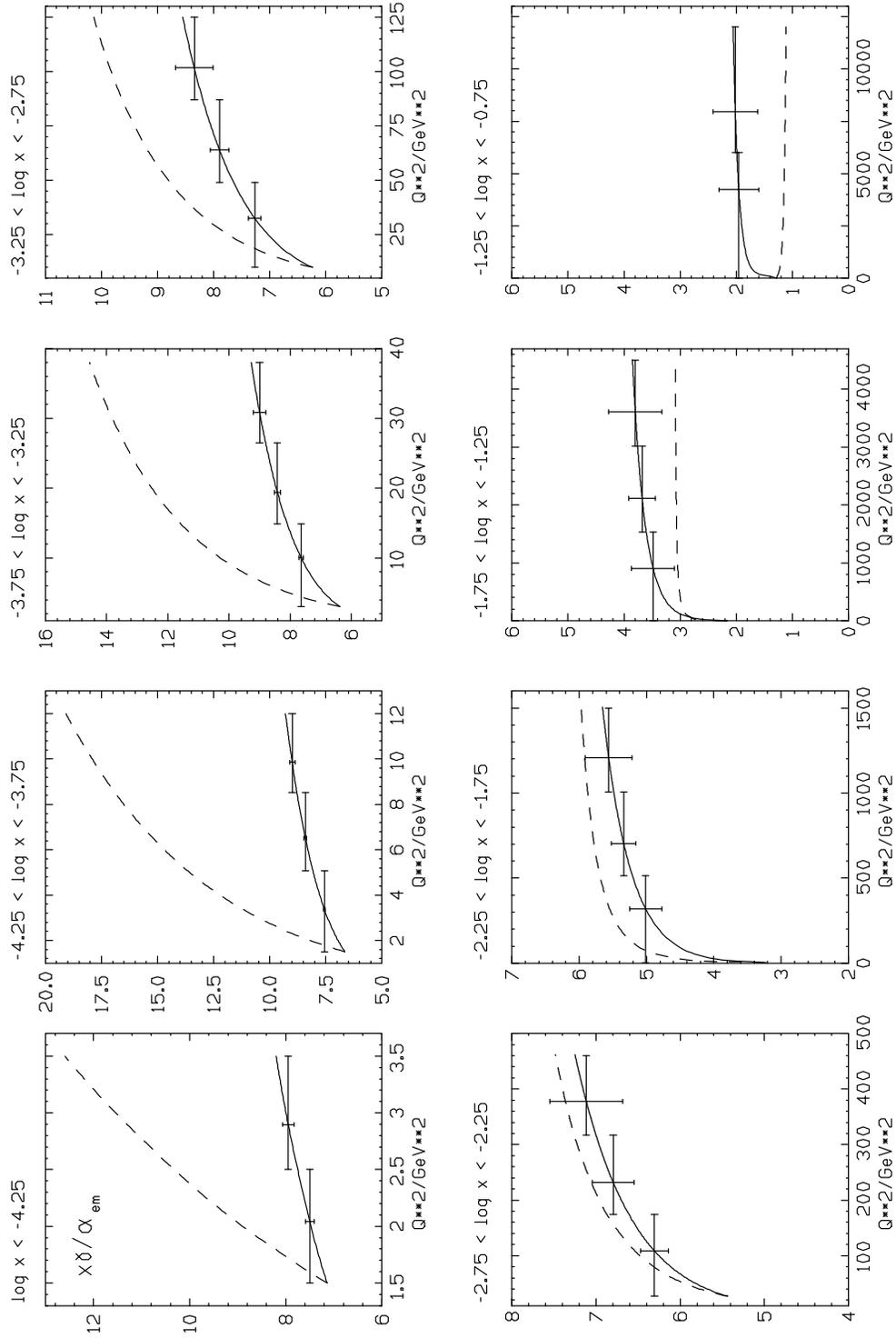}
\vspace*{0.4cm}
\caption{\sf Expected statistical accuracy of a measurement 
of the $Q^2$ dependence of $x\gamma (x,Q^2)/\alpha_{em}$ in various 
$x$-bins. The dashed line displays the $Q^2$ dependence of the 
LO GRV~\cite{grv} gluon density, which has been normalized in each plot so
as to coincide with $x\gamma (x,Q^2)/\alpha_{em}$ at the lowest 
accessible $Q^2$ value.}
\vspace*{-0.5cm}
\end{center}
\end{figure}
The gluonic function $g$
evolves much more strongly at small $x$
than its photon counterpart $\gamma$. 
Only at $x > 10^{-2}$ does the $Q^2$-evolution of $\gamma$ overtake. 
The statistics appear to be insufficient to explore in detail the 
region $x \gtrsim 0.07$, where $g$ starts to decrease 
with $Q^2$, while, for $\gamma$, Eq.~(\ref{evol}) predicts an increase 
for all $x$.

\section {Conclusions}

We have shown that it is possible to extract $\gamma(x,Q^2)$ 
--the function describing the photon constituency of protons--
from existing data. Even its $Q^2$-dependence should be already
observable, allowing for an interesting test of a combination
of QED and QCD. The photon
and gluon distribution functions should
show a strikingly different $Q^2$ evolution.

We know so little about the photonic structure function $\gamma$
that the possible improvements on its theoretical understanding
would be premature, since the extra precision they would bring about
is surely at a more refined level than the current uncertainty. 
The corrections to next-to-leading order in $\alpha_s$ would be
the most interesting, as they should help control the dependence
of the predictions on the choice of variable representing the
momentum scale. As measurements
of DICS materialize --and we hope they soon will--
these improvements should become timely.

\section*{Acknowledgements}
We are grateful to A.\ de Roeck for information and many helpful 
discussions on experimental aspects of the QED Compton process at 
HERA. We thank A.\ Vogt for providing the Fortran code for the parton 
densities of~\cite{grv}. We also acknowledge useful discussions
with G. Altarelli,  D.\ de Florian 
and M.\ Stratmann.
%

\end{document}